# Analysis of temporal characteristics of the editorial processing in scientific periodicals

Olesya Mryglod, [1] *Institute for Condensed Matter Physics of the National Academy of Sciences of Ukraine*;
[2] *Lviv Polytechnic National University*
Yurij Holovatch, *Institute for Condensed Matter Physics of the National Academy of Sciences of Ukraine*
Ihor Mryglod, *Institute for Condensed Matter Physics of the National Academy of Sciences of Ukraine*

**Abstract**

The first part of our work is connected with the analysis of typical random variables for the specific human-initiated process. We study the data characterizing editorial work with received manuscripts in several scientific journals. In such a way we found the waiting time distributions that could be called the typical for an ordinary peer-review scientific journal. In the second part of this study a model of editorial processing of received manuscripts is developed. Within the model, different scenarios of the manuscript editorial processing are examined. Combining the results of the quantitative experiment and model simulations we arrive to the set of conclusions about time characteristics of editorial process in scientific journals and a peer-review contribution.

## 1. Introduction

Nowadays new possibilities for quantitative studying of human activity processes appear. The modern computer technologies allow to collect and to process large volumes of statistical data [1–4]. It is possible to fix the time of each performed operation from large flow of human (customer / user / client) actions and to obtain the data set representing whole picture. Therefore, new methods for data analysis and useful knowledge discovering could be applied. So, in addition to psychological, social and other aspects of human behavior today it is possible to consider it from new point of view, to study it numerically.

The sequences of human actions (telephone calls, information queries or stock exchanges) are not new subject to study. The classical models of human dynamics, used from risk assessment to communications, assume that human actions are randomly distributed in time and thus well approximated by Poisson processes. But the results of recent investigations made evident the non-Poisson statistics for timing of many human dynamics processes: the long periods of inactivity separated by bursts of intensive activity [2,5].

New researches of human activity processes are based on the possibility to operate with numerous statistical data about different human actions such as browsing the Internet, data downloading, electronic communication, initiating financial transactions etc [1–5]. In such studies two random variables are often considered: the time interval between two consecutive actions by individual (called the interevent time $t_{\text{int}}$) and the so-called waiting time $t_{\text{w}}$, the time a task is waiting for an execution. The power-law

$$P(t) \sim t^{-\alpha} \qquad (1)$$

(or close to it) functional forms of probability distributions of these random variables were found. It is interesting to note, that the exponent that governs the power-law dependences differs for different processes [5].

These results attract interest and provoke to analyze new and new processes involving human actions in different fields of our life. The analysis of time statistics of human activity patterns can be useful for different optimization and control tasks in spheres of mass service, communication, information technologies, resource distribution, etc. Besides, this approach can give a new possibility to understand human behavior and to get its additional quantitative measure. Although the origin of power laws is currently under discussion [7, 8] it becomes clear that such dependences describe a lot of natural human activity processes.

## 2. The problem statement

We consider the processing of papers submittion to scientific journals as an example of human activity processes [6]. In this kind of mass service system the input flow consists of submitted papers forming the queue. A standard procedure that follows a paper submission can include following steps: (i) work of referees, (ii) corrections if necessary, (iii) acceptance by an Editorial Board, (iv) other intermediate processes. On each of the stages mentioned above the paper may be rejected. However, typically the information concerning the rejected papers is not publicly available. Therefore, we consider the random variable $t_w$ (waiting time) defined as a number of days between the dates of the paper final acceptance $t_a$ and the paper submission $t_s$:

$$t_w = t_a - t_s. \qquad (2)$$

All the stages of the editorial editing of scientific papers are considered together as the one process (Fig. 1). Though more than one actor takes part in it, we consider that every part of this work is controlled by an Editorial Board. That's why we can treat this process as one of the main characteristics of the Editorial Board work.

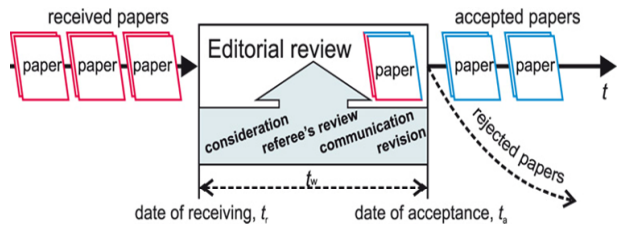

**Fig.1. Schematic picture of editorial processing of papers in scientific journals.**

Our goal in the first part of this study is to determine the functional form of probability distribution $P(t_w)$ based on the statistical data analysis performed for a few scientific journals. Another task is to determine if possible any typical form of $P(t_w)$ for normally working Editorial Board. In the second part of this study we build the model of the editorial processing of received manuscripts. Our purpose here is to make comparison between the empirical data and the modeling results.

## 3. Statistical data analysis

Three scientific editions of the international Elsevier Publishing House with different Editorial Boards were chosen in our study: "Physica A: Statistical Mechanics and its Applications", "Physica B: Condensed Matter" [6] and "Information Systems". The publicly accessible information on official web-site[1] about date of reception and acceptance for scientific papers has been used to calculate waiting times $t_w$. Some formal parameters of databases created for these journals are shown in Table 1.

**Tab. 1**
**Characteristics of analysed databases.**

|  | "Physica A" | "Physica B" | "Information Systems" |
|---|---|---|---|
| Number of records | 4775 | 4944 | 814 |
| Maximal value of $t_w$, days | 1629 | 1087 | 2260 |

---
[1] http://www.elsevier.com

At the first stage the probability histograms of waiting times $P(t_w)$ for selected journals were constructed [6]. Our purpose was to verify the functional form of $P(t_w)$ and to refer it to power-law-like class (non-Poisson processes) or, for example, to exponential-like class (Poisson processes). The exponential distributions of random variables $t_{int}$ and $t_w$ testify to random selection of tasks to execute [7, 8].

We have verified two main hypotheses about form of waiting times probability distributions $P(t_w)$ which describe human activity processes: (i) log-normal distribution [4]:

$$P(t_w) = P_0 + \frac{A}{\sqrt{2\pi}\,\varpi\,t_w} e^{-\frac{\left[\ln\left(\frac{t_w}{t_c}\right)\right]^2}{2\varpi^2}}, \quad t_c, \varpi > 0 \quad (3)$$

where $\ln t_c$ and $\varpi$ are the mean and standard deviations of the $\ln(t_w)$, $P_0, A$ are fitting constants; and (ii) power-law distribution with exponential cutoff [5] for exponent values $\alpha = \{1; 3/2\}$:

$$P(t_w) = A t_w^{-\alpha} e^{-\frac{t_w}{t_0}}, \qquad t_0 > 0, \quad (4)$$

where $t_0$ is characteristic of waiting time which depends on traffic intensity, $A$ is a constant.

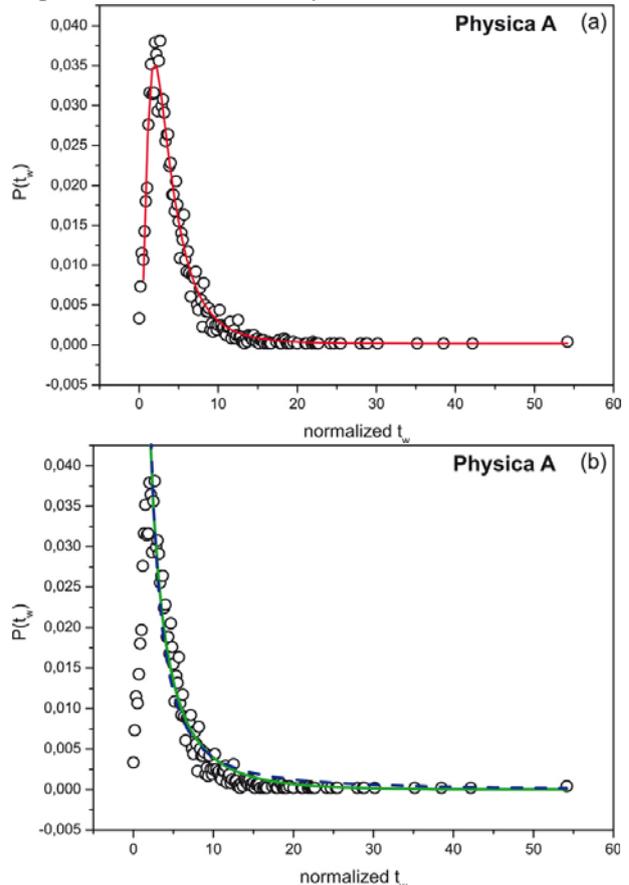

**Fig.2. The $P(t_w)$ distribution for "Physica A" journal with different approximation curves: (a) log-normal (3), (b) power-law with exponential cutoff (4) with $\alpha = 1$ (solid lighter line) and $\alpha = 3/2$ (dashed line).**

Using fitting procedure we found optimal parameters for both distributions (3) and (4). The results obtained for **"Physica A"** journal are shown in Fig. **2**.

We have found that both log-normal and power-law function with exponential cutoff and exponent $a = 1$ can be the probable functions of the $P(t_w)$ distributions. The same conclusions have been done for all three journals.

### 4. Model

We came to conclusion about the typical form of waiting time distributions $P(t_w)$ for scientific journals with normally working Editorial Boards based on results described above. On the one hand, we can say about their closeness to power-law function (1) with exponent $a \approx 1$. On the other hand, they are very close to exponential dependence also. But we can clearly see the general form of obtained $P(t_w)$ distributions with one distinct maximum and the comparatively smooth slope down to the large values of $t_w$. One can also note that the origin of such functional form of $P(t_w)$ distributions is unknown. We can suppose that the contribution of human dynamics could be the reason of observed affinity of $P(t_w)$ to power-laws. The peer-reviewing stage (including the work of authors and the communication process) is the human activity probably most similar to "natural" in comparison with the rest stages of editorial papers processing. The other phases consist of different periodical tasks (Editorial Board meetings, uploads to web-cite, etc.) or work with manuscripts in order of receiving (i. e., language and technical editing).

To verify the hypothesis about the key role of peer-reviewing in the waiting times distributions $P(t_w)$ shaping we built the simple simulation model of editorial work in scientific journal omitting the peer-reviewing [9].

In this case the decisions about all received manuscripts are taken during the subsequent Editorial Board meeting. In our model the period between these sessions is equal to the period of journal publishing. Now the values of waiting time $t_w$ for each published paper could be calculated as a number of days between the moment of its reception and the subsequent Editorial Board meeting (Fig. 3).

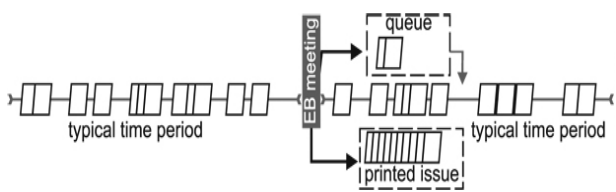

**Fig.3. The schematic representation of the modelled editorial work in scientific journals.**

In the absence of peer-review all the received manuscripts are supposed to be considered (accepted or rejected) by the Editorial Board. In this case the presence of some constrains is important, for example, the limited physical size of issue. For simplicity all the papers in our model have the equal number of pages, so the volume of journal issues could be limited by number of papers (we specify the issue volume equals 10 papers). The value of parameter called the traffic intensity $\rho = \lambda/\mu$ ($\lambda$ is the tasks arrival rate and $\mu$ is the execution rate, respectively) allows to distinguish three regimes of work [5]: (i) subcri­tical, $\rho < 1$; (ii) critical, $\rho = 1$; and (iii) supercritical, $\rho > 1$. The first regime of editorial work is not effective due to the incompleteness of the majority of issues. The third regime is not appropriate also due to the endless queue of papers and therefore increasingly large values of their waiting times. The second regime could be considered as a perfect one since the issues are complete and the papers are published without delays. But this regime is not stable and it can not be reached in practice because it is impossible to control the number of input manuscripts from different authors and thus the value of $\lambda$. It is possible to reach the "pseudocritical" regime of editorial work applying some limitations, for example limitation on queue length. In this situation the queue of papers periodically appears but does not grow infinitely.

Gradually complicating the model we simulate such situation: input flow of manuscripts is modeled as the classical Poisson process; there are two possible ways to choose the papers from the queue: "FIFO" (first-in-first-out) and "RANDOM"; there are also two scenarios to reject the excessive papers limiting the queue length: "LIFO" (last-in-first-out) and "RANDOM"; $\lambda \approx \mu$. Some of the results are presented in Figs. 4, 5. In fact, the principal shape of $P(t_w)$ distribution is determined by a method to choose papers from the queue.

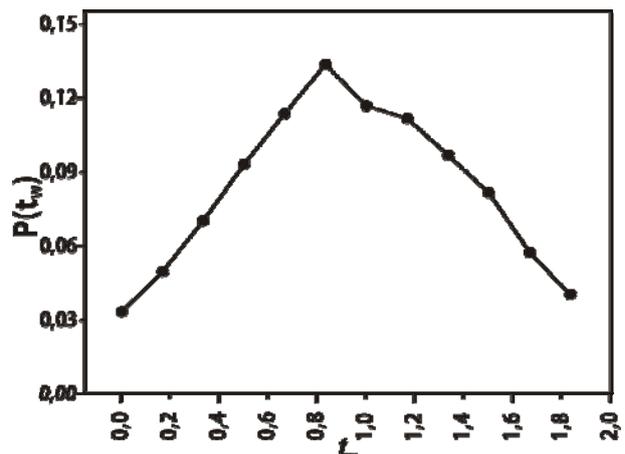

**Fig.4. Probability distribution of waiting times. The results of modeling are presented for: $\lambda = \mu = 10$, the "FIFO" method of papers choosing and the "LIFO" method of their rejection.**

As we can see from Fig. 4, the obtained functional form of $P(t_w)$ distribution considerably differs from the experimental one. The same result was obtained in the case of the "FIFO" method of papers choosing and the "RANDOM" method of their rejection.

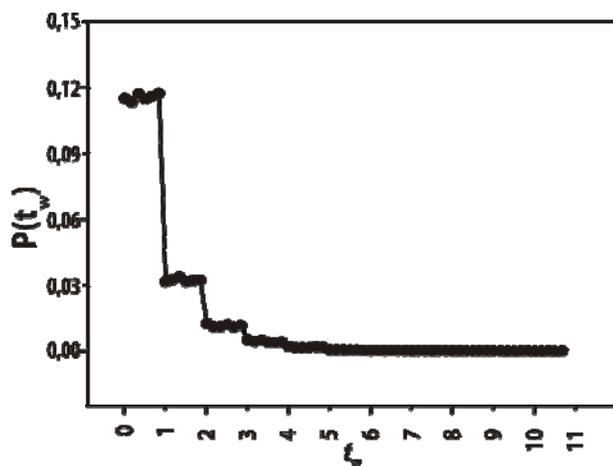

**Fig.5. Probability distribution of waiting times. The results of modeling are presented for:** $\lambda = \mu = 10$, **the "RANDOM" method of papers choosing and the "RANDOM" method of their rejection.**

In Fig. 5 we can see the typical modeling results for the case of "RANDOM" selection of papers from the queue. Here several shelves corresponding to each time period could be observed. Such shape of $P(t_w)$ distribution is the same for both "LIFO" and "RANDOM" methods of papers rejection.

## 5. Conclusions

We have found that both log-normal and power-law function with exponential cutoff and exponent $a = 1$ can be the probable functions of waiting time distributions $P(t_w)$ for manuscripts in scientific journals. In fact, both log-normal and power-law functions predict exactly the same leading power behavior $t^{-1}$, differing only in the functional form of the exponential correction [8]. In this sense, process considered here is governed by similar probability distributions as other examples of human activities [1–5, 7]. The observed data fluctuations can be explained by relatively small statistics but such situation is usual for the majority of real-world data bases. Thus, we consider the obtained form of probability distributions $P(t_w)$ as the typical one that can be used for scientometrical analysis of editions. The length of waiting times is an important characteristic of Editorial Board's work. The publication delay effects journal rankings according to the impact factor as well as personal citation rating of authors.

The simple model of Editorial Board work was created to verify the hypothesis about the leading role of peer-reviewing in the waiting times distributions $P(t_w)$ shaping. In general, the obtained results can be considered as the support of our previous conclusions.

**Authors:**

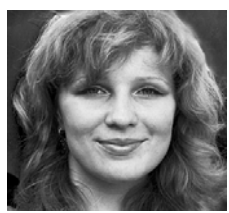

**Olesya Mryglod**
[1] Institute for Condensed Matter Physics of the Nat. Acad. of Sci. of Ukraine,
1 Svientsitskii Str., 79011 Lviv, Ukraine;
[2] Lviv Polytechnic National University,
12 Bandery Str., 79013 Lviv, Ukraine
tel. +38(098)9448326
email: *olesya.m@gmail.com*

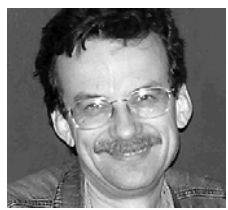

**Yurij Holovatch**
Institute for Condensed Matter Physics of the Nat. Acad. of Sci. of Ukraine,
1 Svientsitskii Str., 79011 Lviv, Ukraine

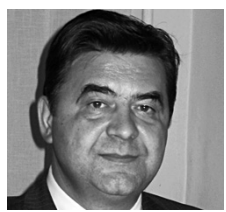

**Ihor Mryglod**
Institute for Condensed Matter Physics of the Nat. Acad. of Sci. of Ukraine,
1 Svientsitskii Str., 79011 Lviv, Ukraine